\documentstyle[preprint,prd,aps,eqsecnum]{revtex}
\begin{document}
\preprint{DART-HEP-93/06\hspace{-26.6mm}\raisebox{-2.4ex}{November 1993}}

\title{Microphysical Approach to Nonequilibrium Dynamics of Quantum Fields}

\author{Marcelo Gleiser\thanks{E-mail: gleiser@peterpan.dartmouth.edu} and
Rudnei O. Ramos\thanks{E-mail: rudnei@northstar.dartmouth.edu}}

\address{{\it Department of Physics and Astronomy, Dartmouth College}\\
{\it Hanover, NH 03755, USA}}

\maketitle

\begin{abstract}
We examine the nonequilibrium dynamics of a self-interacting $\lambda\phi^4$
scalar field theory. Using a real time formulation of finite temperature
field theory we derive, up to two loops and $O(\lambda^2)$, the effective
equation of motion describing the approach to equilibrium. We present
a detailed analysis of the approximations used in order to obtain a
Langevin-like equation of motion, in which the noise and dissipation
terms associated with quantum fluctuations obey a fluctuation-dissipation
relation. We show that, in general, the noise is colored (time-dependent)
and multiplicative (couples nonlinearly to the field), even though it is
still Gaussian distributed. The noise becomes white in the infinite
temperature limit. We also address the effect of couplings to
other fields, which we assume play the r\^ole of the thermal bath,
in the effective equation of motion for $\phi$. In particular, we obtain the
fluctuation and noise terms
due to a quadratic coupling to another scalar field.
\vspace{0.5cm}

\noindent

\end{abstract}

\section{\bf Introduction}

The possibility that the Universe went through a series of phase transitions
as it expanded and cooled from times close to the Planck scale
has been actively investigated for the past
fifteen years or so \cite{kolb}.
It is hoped that by studying the non-trivial dynamics
typical of the approach to equilibrium in complex systems, many of the
current questions of cosmology, from the origin of the baryonic matter excess
to the large-scale structure of the Universe, will be answered in the near
future. As is well-known, the origin of density perturbations that seed
structure formation has been linked to either the existence of topological
defects, such as strings or textures formed during a GUT-scale
transition \cite{defects},
or to inflation in one of its incarnations. In particular, the old, new,
extended, and natural models of inflation all invoke a symmetry breaking
transition in which nonequilibrium conditions play a crucial
r\^ole \cite{inflation}.
At the electroweak scale, the focus has been in generating the
baryon number excess during a first-order phase transition \cite{ew}. Even
though there are certain questions related to the reliability of the
perturbative expansion for weak enough transitions \cite{ewpert}
as well as to the mechanism
by which weak first-order transitions complete \cite{bubbles}, it is
currently believed that nonequilibrium conditions are a crucial ingredient
for baryogenesis.

Despite its relevance, not much has been done to understand nonequilibrium
aspects of phase transitions
in cosmology. (This situation is rapidly changing.
We will soon refer to past and recent work on the subject.)
Most of what has been done so far is related to the finite
temperature effective potential (computed in general to one-loop order)
which, by its very definition, is only adequate to describe equilibrium
situations; the calculation is usually done in euclidean time so that we
can obtain the equilibrium partition function from a transition amplitude.
The great advantage of using the effective potential is that it gives us
information about static properties of the system such as its
possible stable and metastable equilibrium states, and critical temperatures
for phase transitions.
The disadvantage is that we lose all information about
real-time processes, which are crucial to understand the mechanism
by which the system
approaches equilibrium. In fact, the one-loop effective potential does
carry, in a somewhat indirect way, information about unstable states in the
system. These are states which are in the ``spinodal'' region, where the
effective potential is concave. If we start with the system in thermal
equilibrium above the critical temperature and then quench it to below
the critical temperature so that its order parameter takes a value within the
spinodal, the approach to equilibrium will be initially
dominated by the growth of
small amplitude long wavelength fluctuations, in the mechanism known as
spinodal decomposition. Thus, the effective potential tells us that
some states will be unstable, and that their final equilibrium state is at
its global minimum, but it does not tell us how the system gets there.
The reader is referred to the recent work of
Boyanovsky, Lee, and Singh for details \cite{boya2}.

These limitations of the effective potential were pointed out by
Mazenko, Unruh, and Wald, in work where they argued that for strong
enough couplings, the slow-roll approximation necessary for successful
inflation may not be adequate. Instead, the approach to equilibrium
would proceed by the formation and growth of domains, typical of spinodal
decomposition \cite{mazenko}. It was subsequently shown within the context of
the new inflationary model, by both
analytical \cite{albrecht} and phenomenological numerical
methods \cite{feldman} that due to the small
couplings needed for the generation
of density
fluctuations, the slow-roll picture of inflation was correct.

This discussion of the validity of the slow-roll approximation in inflation
raises some very interesting questions related to the way we picture the
approach to equilibrium in field theories, which are quite independent of
inflation. For example, the
distinction between the ``system'', which is out of equilibrium,
and the ``thermal bath'', which drives the system into equilibrium, is somewhat
blurred in the context of nonlinear field theories. In fact, for
self-interacting field theories, the short wavelength modes can serve as the
thermal bath driving the longer wavelength modes,
which have slower dynamics, into
equilibrium. In this sense, the field can be its own thermal bath. Of course,
other fields coupled to the order parameter scalar field (henceforth
the ``system'') may serve as
the  thermal (or, at $T=0$, quantum) bath.

In one of the original works on this
subject which was motivated by cosmology, Hosoya and Sakagami obtained
an approximate dissipation term in the equation of motion
satisfied by the thermal average of
the scalar field, by invoking a small deviation from equilibrium in the
Boltzmann equation for the number density operator. This calculation was
then supplemented by a computation of transport coefficients
using Zubarev's method for nonequilibrium statistical operators \cite{hosoya1}.
Using an approach which is closer to the one we will adopt here,
Morikawa obtained the effective Langevin-like equation (that is, with both
fluctuation and dissipation terms but not quite as simple as the Langevin
equation) for a scalar field
interacting with a fermionic bath using real-time
field-theoretical techniques
at zero and, very briefly, finite temperature \cite{morikawa}.
More recently, Hu, Paz, and Zhang analyzed the case of a quantum bath
given by a scalar
field quadratically coupled to the system
\cite{hu1}, while
Lee and Boyanovsky considered the case of a thermal bath given by a scalar
field linearly coupled to
the system \cite{boya1}. Some works dealing with nonequilibrium
evolution within a cosmological framework can be found in Ref. \cite{cosmos}.
Here we will only be concerned with dynamics in Minkowski spacetime.

Recently, and in particular in Refs. \cite{hu1} and
\cite{boya1},
the properties of the noise as being in general
colored and multiplicative (unless
the coupling between system and bath is linear) have been
emphasized. This can have very important consequences to our understanding
of phase transitions, as suggested by Habib, even though results
at this point are preliminary \cite{habib}. The reason is that potentially,
a multiplicative noise may sharply decrease the relaxation time-scales
in the system and thus accelerate the approach to equilibrium. Numerical
simulations of the approach to equilibrium
have so far employed a phenomenological Langevin equation,
with white and additive noise to mimic
the effects of the thermal bath. In (1+1)-dimensions both the thermal
nucleation of kink-antikink pairs \cite{kinks} and the decay of metastable
states \cite{1ddecay} were studied, while in (2+1)-dimensions the decay
of metastable states was recently investigated \cite{2ddecay}. The time-scales
measured in these simulations agree with the theoretical prediction for the
decay rate, $\Gamma \sim {\rm exp}[-B(T)/T]$, as long as $B(T)$ is
the classical ({\it i.e.}, obtained with the classical potential) nucleation
barrier given by the energy of the appropriate field configuration that
saturates the path integral, the mass of the kink-antikink pair or
the energy of the bounce configuration in the examples mentioned above.

The question then is if the phenomenological Langevin equation used in the
above simulations is indeed reproducing the essential physics of the
approach to equilibrium, or if we are dangerously oversimplifying things.
The above discussion suggests that the effective equation which
describes the approach to equilibrium of the slower
moving modes can
be quite different from the phenomenological Langevin equation with its
white and additive noise. Two tasks are at hand then. First we must obtain
the effective equation for a self-interacting scalar field which acts as
its own bath and compare it with the equation obtained by having another
field act as the bath. This should elucidate the nature of the thermal
bath in these two situations, and also give us an answer as to whether the
phenomenological Langevin equation is at all valid in some limit. The second
task follows naturally the first. Once we have an effective equation we
trust (in some limit), we should use it to simulate numerically
the nonequilibrium
dynamics, measure the relaxation time-scales, and
compare the results with the results obtained with the simplified
phenomenological Langevin equation.

In this paper we will concentrate on the first task. Namely, we will
obtain, within perturbation theory, the effective equation of motion
describing the approach to equilibrium of a self-coupled scalar field.
We will integrate out the short wavelength modes whose influence will be
felt as a thermal bath through the nonlinear couplings to the longer wavelength
modes, which we take as the system. The separation between bath and system is
implemented by perturbation theory, since the effective action is
obtained by {\it integrating} over small fluctuations about the state we
are expanding about. We will include corrections up to two-loops, as
nonvanishing viscosity (and transport coefficients, in general) terms
in finite temperature field theory only show up by considering higher
order (loop) corrections to the field propagators and are dependent on the
imaginary part (decay width) of the self-energy corrections
\cite{hosoya1,hosoya2,chou,jeon}.
Fluctuation terms are obtained by associating the imaginary terms
in the effective action as coming from the interaction of $\varphi$ with
fluctuating (noise) fields, as done in \cite{morikawa} and \cite{boya1}.
We will also obtain the effective equation of motion
in the presence of another scalar
field quadratically coupled to the system, thus reproducing (even though
we focus more on dynamical aspects) the
analysis of Ref. \cite{hu1} for finite temperatures. Our results could,
for example, be used in the numerical
investigation of symmetry restoration at finite
temperature \cite{abraham}.

The paper is organized as follows: In Sec. II we derive
the effective action for a nonuniform time-dependent background field
configuration $\varphi(\vec x,t)$, up to two loops and order $\lambda^2$.
In Sec. III we obtain the effective equation of motion for
$\varphi(\vec x,t)$ and discuss the approximations involved in order that it
obeys a Langevin-like equation.
In Sec. IV we examine
the effect of other fields interacting with
the scalar field, by studying the case of a quadratically coupled
scalar field and by evaluating its contributions to noise and dissipative
terms.
Conclusions are presented in Sec. V. Two appendices are included in order
to obtain some technical results used in the paper.

\section{The Two-Loop Finite Temperature Effective Action}

Consider the scalar field model with Lagrangian density

\begin{equation}
{\cal L}[\phi] = \frac{1}{2} \left( \partial_\mu \phi \right)^2
-\frac{m^2}{2} \phi^2 - \frac{\lambda}{4 !} \phi^4
\label{Lagr}
\end{equation}

\noindent
and with generating functional $Z[J]$, in terms of an external source $J$,
given by

\begin{equation}
Z[J] = \int_c D \phi \exp \left\{ i S[\phi,J]\right\} \:,
\label{ZJ}
\end{equation}

\noindent
where the classical action is given by

\begin{equation}
S[\phi,J] = \int_c d^4 x \left\{ {\cal L}[\phi] + J(x) \phi(x) \right\} \: .
\label{SphiJ}
\end{equation}

\noindent
In (\ref{SphiJ}) the time integration is along a contour suitable for
real-time evaluations, which we choose as being Schwinger's closed-time
path \cite{schwinger,chou,rivers}, where the time path $c$
goes from $-\infty$ to $+\infty$ and then back to $-\infty$. The functional
integration in (\ref{ZJ}) is over fields along this time contour. As with
the euclidean time formulation, the scalar field is still periodic in time,
but now with $\phi(t, \vec x)=\phi(t-i\beta, \vec x)$. Temperature appears
due to the boundary condition, but now time is explicitly present in the
integration contour.

As usual, the effective action $\Gamma[\varphi]$ is defined in terms of the
connected generating functional $W[J]$ as

\begin{equation}
\Gamma[\varphi] = W[J] - \int_c d^4 x J(x) \varphi(x) \: ,
\label{action}
\end{equation}

\noindent
with $\varphi(x)$ defined by $\varphi(x) \equiv
\frac{\delta W[J]}{\delta J(x)}$,
and

\begin{equation}
W[J] = - i \ln \int_c D \phi \exp\left\{ i S[\phi,J] \right\} \: .
\label{WJ}
\end{equation}

The perturbative loop expansion for $\Gamma[\varphi]$ is obtained
by writing the scalar field as $\phi \to \phi_0 + \eta$,
where $\phi_0$ is a field configuration which extremizes the classical
action $S[\phi,J]$ and $\eta$ is a small perturbation about this configuration.
By using (\ref{action}) and (\ref{WJ}), we can relate $\phi_0$ to $\varphi$
($\phi_0 = \varphi - \eta$) and write the effective action, for $J\to 0$,
at one-loop order, as

\begin{equation}
\Gamma[\varphi] = S[\varphi] + \frac{i}{2} {\rm Tr} \ln \left[
\Box + V''(\varphi)\right]  \:,
\label{action 1l}
\end{equation}

\noindent
where

\begin{equation}
\frac{i}{2} {\rm Tr} \ln \left[\Box + V''(\varphi)\right] =
- i \ln \int_c D \eta \exp \left\{ - \frac{i}{2} \eta \left[
\Box + V''(\varphi) \right] \eta \right\} \: .
\label{Trln}
\end{equation}

Negleting contributions to (\ref{action 1l}) which are independent of
$\varphi$, we can expand the
logarithm in (\ref{action 1l}) as

\begin{eqnarray}
\lefteqn{\frac{i}{2} {\rm Tr} \ln \left[\Box + V''(\varphi)\right] =
\frac{i}{2} {\rm Tr}_c \sum_{m=1}^{+\infty} \frac{(-1)^{m+1}}{m}
G_\phi^m \left( \frac{\lambda}{2}\varphi^2\right)^m=}
\nonumber \\
& & = \frac{i}{2} \sum_{m=1}^{+\infty} \frac{(-1)^{m+1}}{m}
\left(\frac{\lambda}
{2}\right)^m {\rm Tr} \int d^4 x_1 \ldots d^4 x_m
G_\phi^{n_1, l_1}(x_1 - x_2) \left[\varphi^2(x_2)\right]_{l_1, n_2}
G_\phi^{n_2,l_2}(x_2-x_3) \ldots
\nonumber \\
& & \ldots  \left[\varphi^2(x_m)\right]_{l_{m-1},n_m}
G_\phi^{n_m,l_m}(x_m-x_1) \left[\varphi^2(x_1)\right]_{l_m,n_{m+1}}
\: .
\label{Trlnsum}
\end{eqnarray}

\noindent
The matrix representation in (\ref{Trlnsum}) is a consequence of the
time contour, since now we must identify field variables with arguments
on the positive or negative directional branches of the time path,
that we denote by $\varphi_{+}$ and $\varphi_{-}$, respectively.
As a consequence of this doubling of field variables, we also have
that $G_\phi^{n,l}(x-x')$, the real-time free field
propagators on the contour, are given by ($l,n = +,-$)
\cite{chou,weert}

\begin{eqnarray}
\lefteqn{G_\phi^{++}(x-x') = i \langle T_{+} \phi(x) \phi(x')\rangle}
\nonumber \\
& & G_\phi^{--}(x - x') = i \langle T_{-} \phi(x) \phi(x')\rangle
\nonumber \\
& & G_\phi^{+-}(x-x') = i \langle \phi(x') \phi(x)\rangle
\nonumber \\
& & G_\phi^{-+}(x-x') = i \langle \phi(x) \phi(x')\rangle \: ,
\label{twopoint}
\end{eqnarray}

\noindent
where $T_{+}$ and $T_{-}$ indicate chronological and anti-chronological
ordering, respectively.
$G_\phi^{++}$ is the usual physical (causal)
propagator. The other three propagators come as a consequence of the
time contour and are considered as auxiliary (unphysical)
propagators \cite{weert}. The explicit expressions for
$G_\phi^{n,l}(x-x')$ in terms of its momentum space Fourier
transforms are given by \cite{chou,rivers}

\begin{equation}
G_\phi(x-x') = i \int \frac{d^3 k}{(2 \pi)^3} e^{i \vec k . (\vec x - \vec x')}
\left(
\begin{array}{ll}
G_\phi^{++}(\vec{k}, t- t') & \:\: G_\phi^{+-}(\vec k, t-t') \\
G_\phi^{-+}(\vec{k}, t- t') & \:\: G_\phi^{--}(\vec k, t-t')
\end{array}
\right) \: ,
\label{Gmatrix}
\end{equation}

\noindent
where

\begin{eqnarray}
\lefteqn{G_\phi^{++}(\vec k , t-t') = G_\phi^{>}(\vec k,t-t')
\theta(t-t') + G_\phi^{<}(\vec k,t-t') \theta(t'-t)}
\nonumber \\
& & G_\phi^{--}(\vec k , t-t') = G_\phi^{>}(\vec k,t-t')
\theta(t'-t) + G_\phi^{<}(\vec k,t-t') \theta(t-t')
\nonumber \\
& & G_\phi^{+-}(\vec k , t-t') = G_\phi^{>}(\vec k,t-t')
\nonumber \\
& & G_\phi^{-+}(\vec k,t-t') = G_\phi^{<}(\vec k,t-t')
\label{G of k}
\end{eqnarray}

\noindent
and, for free propagators at finite temperature,

\begin{eqnarray}
& G_\phi^{>}(\vec k ,t-t') &= \frac{1}{2 \omega(\vec k)}
\Bigl [ (1 + 2 n(\omega)) \cos[\omega(t-t')] - i \sin[\omega(t-t')]
\Bigr ]
\nonumber \\
& G_\phi^{<}(\vec k , t-t') & = G_\phi^{>}(\vec k, t'-t) \: ,
\label{G><}
\end{eqnarray}

\noindent
where $n(\omega)= \left(e^{\beta \omega} - 1\right)^{-1}$ is the Bose
distribution and $\omega\equiv \omega(\vec k)$ is the free particle energy,
$\omega(\vec k) = \sqrt{\vec{k}^2 + m^2}$.

Let us now add to (\ref{action 1l}) contributions up to two-loops and order
$\lambda^2$. Graphically we have,

\begin{equation}
\begin{picture}(270,28)

\put(-85,0){$\Gamma[\varphi]= S[\varphi] \; + $}

\thicklines

\put(0,0){\line(-1,-1){10}}
\put(10,0){\circle{20}}
\put(0,0){\line(-1,1){10}}

\put(25,0){+}

\put(50,0){\line(-1,-1){10}}
\put(60,0){\circle{20}}
\put(50,0){\line(-1,1){10}}
\put(70,0){\line(1,1){10}}
\put(70,0){\line(1,-1){10}}

\put(85,0){+}

\put(110,0){\line(-1,-1){10}}
\put(110,0){\line(-1,1){10}}
\put(120,0){\circle{20}}
\put(140,0){\circle{20}}

\put(155,0){+}

\put(170,0){\line(1,0){40}}
\put(190,0){\circle{20}}

\put(220,0){$+\: {\cal O}(\lambda^3)\: ,$}

\end{picture}
\label{action 2l}
\end{equation}

\vspace{0.5cm}
\noindent
where, in the graphic representation, $\varphi$ is in the external legs
and the internal propagators are given by $G_\phi^{n,l}$.
In terms of the field variables $\varphi_{+}$ and $\varphi_{-}$, the terms
in Eq. (\ref{action 2l}) are given by (note that now time runs only forward)

\begin{equation}
S[\varphi] = \int d^4 x \left\{ {\cal L}[\varphi_{+}] - {\cal L}[\varphi_{-}]
\right\} \:,
\label{S+-}
\end{equation}

\begin{eqnarray}
\begin{picture}(180,28)

\thicklines

\put(0,0){\line(-1,-1){10}}
\put(10,0){\circle{20}}
\put(0,0){\line(-1,1){10}}
\end{picture}
\hspace{-5.5cm}=
-\frac{\lambda}{4} {\rm Tr} \int d^4 x \int \frac{d^3 q}{(2 \pi)^3}
\left(
\begin{array}{ll}
G_\phi^{++}(\vec{q}, 0) & \:\: G_\phi^{+-}(\vec q, 0) \\
G_\phi^{-+}(\vec{q}, 0) & \:\: G_\phi^{--}(\vec q, 0)
\end{array}
\right)
\left(
\begin{array}{ll}
\varphi_{+}^2 (x) & \:\: 0 \\
0 & \:\: -\varphi_{-}^2 (x)
\end{array}
\right) =
\nonumber
\end{eqnarray}

\begin{equation}
\hspace{-1.5cm}= - \frac{\lambda}{4} \int d^4 x \left[ \varphi_{+}^2 (x) -
\varphi_{-}^2 (x) \right] \int \frac{d^3 q}{(2 \pi)^3}
\frac{1}{2 \omega(\vec q)} \left[ 1 + 2 n(\omega)\right] \: ,
\label{1loop1}
\end{equation}

\noindent
where $G_{\phi}^{n,l}(\vec q ,0)$ is given by (\ref{G of k}) (for $t-t'=0$).
Eq. (\ref{1loop1}) gives just the finite temperature mass contribution to
the effective action (renormalized by a proper mass counterterm $\delta
m^2$ which we are not including here).
The second graph in (\ref{action 2l}) is given by

\begin{eqnarray}
\begin{picture}(180,28)
\thicklines
\put(20,0){\line(-1,-1){10}}
\put(30,0){\circle{20}}
\put(20,0){\line(-1,1){10}}
\put(40,0){\line(1,1){10}}
\put(40,0){\line(1,-1){10}}
\put(60,0){=}
\end{picture}
\hspace{18cm}
\nonumber
\end{eqnarray}

\begin{eqnarray}
&=& i \frac{\lambda^2}{16} {\rm Tr} \int d^4 x d^4 x' \int \frac{d^3 k}{
(2 \pi)^3} e^{i \vec k . (\vec{x}-\vec{x'})} \int \frac{d^3 q}{(2 \pi)^3}
\left(
\begin{array}{ll}
G_\phi^{++}(\vec{q}, t- t') & \:\: G_\phi^{+-}(\vec q, t-t') \\
G_\phi^{-+}(\vec{q}, t- t') & \:\: G_\phi^{--}(\vec q, t-t')
\end{array}
\right) \times \nonumber \\
& \times &
\left(
\begin{array}{cc}
\varphi_{+}^2 (x') & \:\: 0 \\
0 & \:\: -\varphi_{-}^2 (x')
\end{array}
\right)
\left(
\begin{array}{ll}
G_\phi^{++}(\vec q -\vec{k}, t- t') & \:\: G_\phi^{+-}(\vec q -\vec k, t-t') \\
G_\phi^{-+}(\vec q -\vec{k}, t- t') & \:\: G_\phi^{--}(\vec q - \vec k, t-t')
\end{array}
\right)
\left(
\begin{array}{cc}
\varphi_{+}^2 (x) & \:\: 0 \\
0 & \:\: -\varphi_{-}^2 (x)
\end{array}
\right) = \nonumber \\
& = & i \frac{\lambda^2}{16} \int d^4 x d^4 x' \int \frac{d^3 k}{(2 \pi)^3}
e^{i \vec k . (\vec x - \vec{x'})} \int \frac{d^3 q}{(2 \pi)^3}
\left[ \varphi_{+}^2 (x) G_{\phi}^{++}(\vec q , t-t')
G_{\phi}^{++}(\vec q - \vec k , t-t') \varphi_{+}^2 (x') -
\right. \nonumber \\
&-& \left. \varphi_{+}^2 (x) G_{\phi}^{+-} (\vec q , t-t')
G_{\phi}^{+-}(\vec q -\vec k ,t-t') \varphi_{-}^2 (x')
-  \varphi_{-}^2 (x) G_{\phi}^{-+}(\vec q , t-t')
G_{\phi}^{-+}(\vec q -\vec k , t-t') \varphi_{+}^2 (x') +\right.
\nonumber \\
& + & \left.
\varphi_{-}^2 (x) G_{\phi}^{--}(\vec q , t-t')G_{\phi}^{--}(\vec q -\vec k ,
t-t') \varphi_{-}^2 (x') \right] \: .
\label{1loop2}
\end{eqnarray}

\noindent
Equivalently, we get for the third graph in (\ref{action 2l}) the expression

\begin{eqnarray}
\begin{picture}(180,28)
\thicklines

\put(15,0){\line(-1,-1){10}}
\put(15,0){\line(-1,1){10}}
\put(25,0){\circle{20}}
\put(45,0){\circle{20}}
\end{picture}
\hspace{-4cm}= -i \frac{\lambda^2}{8} \int \frac{d^3 k}{(2 \pi)^3}
\frac{\left( 1 + 2 n(\omega)\right)}{2 \omega(\vec k)}
\int d^4 x \int d t' \int
\frac{d^3 q}{(2 \pi)^3} \left\{ \varphi_{+}^2 (x) \left[G_{\phi}^{++}
(\vec q ,t-t') \right]^2 - \right.\nonumber
\end{eqnarray}

\begin{equation}
\left. - \varphi_{+}^2 (x) \left[G_\phi^{+-}(\vec q ,t-t')\right]^2 -
\varphi_{-}^2 (x) \left[G_\phi^{-+}(\vec q ,t-t')\right]^2
+ \varphi_{-}^2 (x) \left[G_\phi^{--}(\vec q ,t-t')\right]^2
\right\}
\label{2loop1}
\end{equation}

\noindent
whereas the fourth graph in (\ref{action 2l}) is given by

\begin{eqnarray}
\begin{picture}(280,28)
\thicklines
\put(0,0){\line(1,0){40}}
\put(20,0){\circle{20}}
\end{picture}
\hspace{-8.2cm}= i \frac{\lambda^2}{12}\int d^4 x d^4 x' \int \frac{d^3 k}{
(2 \pi)^3} e^{i \vec k . (\vec{x} - \vec{x'})} \int
\frac{d^3 q_1}{(2 \pi)^3}\frac{d^3 q_2}{(2 \pi)^3}\frac{d^3 q_3}{(2 \pi)^3}
\delta(\vec k - \vec{q}_1 - \vec{q}_2 - \vec{q}_3) \times \nonumber
\end{eqnarray}
\vspace{-0.8cm}
\begin{eqnarray}
& \times & \left\{ \varphi_{+}(x) G_\phi^{++}(\vec{q}_1 ,t-t')
G_\phi^{++}(\vec{q}_2 ,t-t')G_\phi^{++}(\vec{q}_3 ,t-t')\varphi_{+}(x')
-\right. \nonumber \\
& - & \left. \varphi_{+}(x) G_\phi^{+-}(\vec{q}_1 ,t-t')
G_\phi^{+-}(\vec{q}_2 ,t-t')G_\phi^{+-}(\vec{q}_3 ,t-t')\varphi_{-}(x')-
\right.\nonumber \\
& - & \left.
\varphi_{-}(x) G_\phi^{-+}(\vec{q}_1 ,t-t')
G_\phi^{-+}(\vec{q}_2 ,t-t')G_\phi^{-+}(\vec{q}_3 ,t-t')\varphi_{+}(x')+
\right. \nonumber \\
& + & \left. \varphi_{-}(x) G_\phi^{--}(\vec{q}_1 ,t-t')
G_\phi^{--}(\vec{q}_2 ,t-t')G_\phi^{--}(\vec{q}_3 ,t-t')\varphi_{-}(x')
\right\} \: .
\label{2loop2}
\end{eqnarray}

Before continuing, it is advantageous to rewrite the field variables
$\varphi_{+}$ and $\varphi_{-}$ in (\ref{action 2l}) in terms of
new field variables $\varphi_c$ and $\varphi_{\Delta}$, defined
by

\begin{eqnarray}
& \varphi_{+} & = \frac{1}{2} \varphi_{\Delta} + \varphi_c\: ,\nonumber\\
& \varphi_{-} & = \varphi_c - \frac{1}{2} \varphi_{\Delta} \: .
\label{newfields}
\end{eqnarray}

\noindent
The physical meaning of these variables is suggested
in Ref. \cite{MSR}, with $\varphi_{\Delta}$ being basically
associated with a response field while $\varphi_c$ is the physical field,
which ``feels'' the fluctuations of the system. The change of variables
(\ref{newfields}) will allow us to identify, in the effective action,
the terms responsible for the fluctuations in the system (the imaginary
terms). The association of $\varphi_c$ as the physical field imposes that we
take $\varphi_{\Delta}=0$ ($\varphi_{+} = \varphi_{-}$)
at the end of the calculation \cite{chou,morikawa}. In terms of the
new variables $\varphi_c$ and $\varphi_{\Delta}$, using (\ref{G of k}) and
(\ref{G><}), we get the following expression for the effective action
(\ref{action 2l}), using the physical propagator $G_\phi^{++}(\vec q,t-t')$
in the Feynman diagrams
(\ref{S+-})-(\ref{2loop2}),

\begin{eqnarray}
\Gamma[\varphi_{\Delta},\varphi_c]& =&\int d^4 x \left\{
\varphi_{\Delta}(x) \left[ -\Box -m^2 - \frac{\lambda}{2}
\int \frac{d^3 k}{(2 \pi)^3} \frac{(1 + 2 n(\omega))}{2 \omega(\vec k)}
+\right.\right. \nonumber \\
&+ & \left.\left.  \frac{\lambda^2}{2} \int d t' \int \frac{d^3 q}{(2 \pi)^3}
{\rm Im}\left[G_\phi^{++}(\vec q , t-t')\right]^2 \theta(t - t')
\int \frac{d^3 k}{(2 \pi)^3} \frac{(1 + 2 n(\omega))}{2 \omega(\vec k)}
\right]\varphi_c (x) - \right.\nonumber\\
&- & \left.
\frac{\lambda}{4 !} \left( 4 \varphi_{\Delta}(x)\varphi_c^3 (x) +
\varphi_{\Delta}^3 (x) \varphi_c (x) \right) \right\}
+ \nonumber\\
&+&\int d^4 x d^4 x' \int \frac{d^3 k}{(2 \pi)^3}
e^{i \vec k . (\vec{x}-\vec{x'})}  \left\{
-\frac{\lambda^2}{8} \left[ \varphi_{\Delta}(x) \varphi_c (x)
\varphi_{\Delta}^2 (x') + \right. \right.
\nonumber \\
&+ & \left.\left.  4 \varphi_{\Delta} (x) \varphi_c (x)
\varphi_{c}^{2} (x') \right]
\int \frac{d^3 q}{(2 \pi)^3} {\rm Im} \left[
G_\phi^{++}(\vec q ,t-t') G_\phi^{++}(\vec q -\vec k ,t-t')\right]
\theta(t -t') - \right. \nonumber \\
& -& \left. \frac{\lambda^2}{3} \varphi_{\Delta}(x) \varphi_c (x')
{\rm Im}\left[\prod_{j=1}^3 \int \frac{d^3 q_j}{(2 \pi)^3}
G_\phi^{++}(\vec{q}_j,t-t')\right]
\theta(t-t') \delta(\vec k - \vec{q}_1 - \vec{q}_2 - \vec{q}_3) +
\right. \nonumber \\
&+ & \left. i \frac{\lambda^2}{4}  \varphi_{\Delta} (x) \varphi_c (x)
\varphi_{\Delta} (x') \varphi_c (x') {\rm Re} \int \frac{d^3 q}{(2 \pi)^3}
\left[G_\phi^{++}(\vec q ,t-t') G_\phi^{++}(\vec q -\vec k ,t-t')
\right] + \right. \nonumber \\
&+ & \left. i \frac{\lambda^2}{12} \varphi_{\Delta} (x) \varphi_{\Delta} (x')
{\rm Re} \left[ \prod_{j=1}^3 \int \frac{d^3 q_j}{(2 \pi)^3}
G_\phi^{++}(\vec{q}_j,t-t')\right]
\delta(\vec k - \vec{q}_1 - \vec{q}_2 - \vec{q}_3) \right\} \: .
\label{full action}
\end{eqnarray}

\noindent
The last two terms in $\Gamma[\varphi_{\Delta}\varphi_c]$, Eq.
(\ref{full action}), give the imaginary contributions to the effective
action at the order of perturbation theory considered. It is straightforward to
associate the imaginary terms in (\ref{full action}) as coming from
functional integrations over Gaussian fluctuation fields $\xi_1$ and
$\xi_2$ \cite{morikawa,boya1}

\begin{eqnarray}
\int D\xi_1 P[\xi_1] \int D \xi_2 & P[\xi_2] & \exp\left\{i \int
d^4 x \left[ \varphi_{\Delta} (x) \varphi_c (x) \xi_1 (x) +
\varphi_{\Delta} (x) \xi_2 (x) \right] \right\} = \nonumber \\
&=& \exp\left\{i \int d^4 x d^4 x' \left[ i \frac{\lambda^2}{4}
\varphi_{\Delta} (x) \varphi_c (x) {\rm Re}\left[G_\phi^{++}\right]_{x,x'}^2
\varphi_{\Delta} (x') \varphi_c (x') +\right. \right.\nonumber \\
& + &  \left.\left.
i \frac{\lambda^2}{12} \varphi_{\Delta}(x){\rm Re}\left[G_\phi^{++}
\right]_{x,x'}^3 \varphi_{\Delta}(x') \right]\right\}\: ,
\label{gaussians}
\end{eqnarray}

\noindent
where $P[\xi_1]$ and $P[\xi_2]$, the probability distributions for
$\xi_1$ and $\xi_2$, respectively, are given by

\begin{equation}
P[\xi_1] = N_1^{-1} \exp\left\{- \frac{1}{2} \int d^4 x d^4 x' \xi_1 (x)
\left( \frac{\lambda^2}{2} {\rm Re} \left[ G_\phi^{++}\right]_{x,x'}^2
\right)^{-1} \xi_1 (x') \right\} \: ,
\label{P1}
\end{equation}

\begin{equation}
P[\xi_2]= N_2^{-1} \exp\left\{- \frac{1}{2} \int d^4 x d^4 x' \xi_2 (x)
\left( \frac{\lambda^2}{6} {\rm Re} \left[ G_\phi^{++}\right]_{x,x'}^3
\right)^{-1} \xi_2 (x') \right\} \:,
\label{P2}
\end{equation}

\noindent
where $N_1^{-1}$ and $N_2^{-1}$ are normalization factors, and
in (\ref{gaussians})-(\ref{P2}) we introduced the compact notation,

\begin{equation}
\left[G_\phi^{++}\right]_{x,x'}^2 = \int \frac{d^3 k}{(2 \pi)^3}
\exp\left [ i \vec k . (\vec x - \vec{x'})\right ] \int \frac{d^3 q}{(2 \pi)^3}
G_\phi^{++}(\vec q ,t-t')G_\phi^{++}(\vec q - \vec k,t-t')
\label{G2}
\end{equation}

\noindent
and

\begin{equation}
\left[G_\phi^{++}\right]_{x,x'}^3 = \int \frac{d^3 k}{(2 \pi)^3}
\exp\left [i \vec k . (\vec x - \vec{x'})\right ] \left [\prod_{j=1}^3
\int \frac{d^3 q_j}{(2 \pi)^3} G_\phi^{++}(\vec q_j,t-t')
\right] \delta(\vec k - \vec q_1 - \vec q_2 - \vec q_3)\: .
\label{G3}
\end{equation}

\noindent
Therefore, using (\ref{gaussians}), Eq. (\ref{full action}) can be
rewritten as

\begin{equation}
\Gamma[\varphi_{\Delta},\varphi_c] = \frac{1}{i} \ln \int D \xi_1
P[\xi_1] \int D \xi_2 P[\xi_2] \exp\left\{ i S_{\rm eff}[\varphi_{\Delta},
\varphi_c,\xi_1,\xi_2]\right\} \: ,
\label{fluc action}
\end{equation}

\noindent
where

\begin{equation}
S_{\rm eff}[\varphi_{\Delta},\varphi_c,\xi_1,\xi_2] =
{\rm Re} \Gamma[\varphi_{\Delta},\varphi_c] + \int d^4 x \left[
\varphi_{\Delta}(x) \varphi_c (x) \xi_1 (x) + \varphi_{\Delta} (x) \xi_2 (x)
\right]\: ,
\label{Seff}
\end{equation}

\noindent
and ${\rm Re} \Gamma[\varphi_{\Delta},\varphi_c]$ is the real part of
Eq. (\ref{full action}). In (\ref{Seff}), the fields $\xi_1$ and $\xi_2$,
with probability distributions given by (\ref{P1}) and (\ref{P2}),
respectively,
act as fluctuation sources for the scalar field configuration $\varphi$.
$\xi_1$ couples with both the response field $\varphi_{\Delta}$ and with
the physical field $\varphi_c$, leading to a coupled (multiplicative) noise
term ($\varphi_c \xi_1$) in the equation of motion for $\varphi_c$,
while $\xi_2$ gives origin to an additive noise term.
In the next section we examine the
relevance of each of these noise terms in the equation
of motion for the physical field $\varphi_c$ and evaluate the
dissipation coefficients associated with them.

\section{The Effective Equation of Motion}

The equation of motion for $\varphi_c$ is defined by

\begin{equation}
\frac{\delta S_{\rm eff}[\varphi_{\Delta},\varphi_c,\xi_1,\xi_2]}
{\delta \varphi_{\Delta}}|_{\varphi_{\Delta} = 0} = 0 ~~.
\label{motion}
\end{equation}

\noindent
Using (\ref{Seff}) and (\ref{full action}), we  obtain

\begin{eqnarray}
\lefteqn{\left[ \Box + m^2 + \frac{\lambda}{2} \int \frac{d^3 k}{(2 \pi)^3}
\frac{1 + 2 n(\omega)}{2 \omega(\vec k)} \left( 1 -
\lambda \int_{-\infty}^{t} d t' \int \frac{d^3 q}{(2 \pi)^3} {\rm Im}
\left[G_\phi^{++}(\vec q , t-t')\right]^2 \right)\right]\varphi_c (x) +
\frac{\lambda}{3 !} \varphi_c^3 (x) }\nonumber \\
& & + \frac{\lambda^2}{2} \varphi_c (x) \int d^3 x' \int_{-\infty}^{t}
d t' \varphi_{c}^2 (\vec{x'},t') {\rm Im} \left[G_\phi^{++}\right]_{x,x'}^2
+ \frac{\lambda^2}{3} \int d^3 x' \int_{-\infty}^{t} d t'
\varphi_c (\vec{x'},t') {\rm Im}\left[G_\phi^{++}\right]_{x,x'}^3 =
\nonumber \\
& & = \varphi_c (x) \xi_1 (x) + \xi_2 (x) \: ,
\label{full eqmotion}
\end{eqnarray}

\noindent
where $\left[G_\phi^{++}\right]_{x,x'}^2$ and $\left[G_\phi^{++}
\right]_{x,x'}^3$ are given by (\ref{G2}) and (\ref{G3}), respectively.
In order to obtain a Langevin-like equation, a series of approximations
must be performed in the above equation of motion. These approximations will
certainly limit the scope of applicability of the final equation to be
obtained (very much as in linear response theory), but on the other hand will
elucidate important aspects of the nonequilibrium physics.
Strictly speaking, a Langevin-like equation can only
be used to describe the nonequilibrium dynamics
of slowly varying modes in near-equilibrium
situations.
To see this, we now focus on the last two terms on the
left hand side of Eq. (\ref{full eqmotion}).

\subsection{Dissipation Coefficients}

Let us first consider the term in the equation of motion dependent on
$\left[G_\phi^{++}\right]_{x,x'}^2$. Inspecting (\ref{G2}),
it is clear that the spatial nonlocality can be handled by
considering only contributions
with zero external momentum, as in the computation of linear response
functions \cite{jeon}. This is what is usually done in the
computation of the one-loop
effective potential as an expansion of vertex functions with
zero external momentum, which is physically
equivalent to considering only nearly spatially homogeneous
fields. We thus obtain,

\begin{eqnarray}
\lefteqn{\frac{\lambda^2}{2} \varphi_c (x) \int d^3 x' \int_{-\infty}^t d t'
\varphi_c^2 (\vec{x'},t') {\rm Im} \left[G_\phi^{++}\right]_{x,x'}^2 =}
\hspace{4cm}\nonumber \\
& & =\frac{\lambda^2}{2}\varphi_c (\vec x,t)
 \int_{-\infty}^t d t' \left[ \varphi_c^2
(\vec x , t') -
\varphi_c^2 (\vec x, t) \right] \int \frac{d^3 q}{(2\pi)^3} {\rm Im}
\left[G_\phi^{++}(\vec q , t-t')\right]^2 +\nonumber \\
& & +\frac{\lambda^2}{2} \varphi_c^3 (\vec x, t) \int_{-\infty}^t d t'
\int \frac{d^3 q}{(2\pi)^3} {\rm Im}\left[G_\phi^{++}(\vec q , t-t')\right]^2
\: ,
\label{Diss 1}
\end{eqnarray}

\noindent
where we have summed and subtracted in (\ref{Diss 1}) the last term
in the rhs. In order to handle the temporal nonlocality let us further
assume that $\varphi_c$ varies sufficiently slowly in time, so that
we  can expand the first term in the rhs of
(\ref{Diss 1}) to first order around $t$. This is
a valid assumption for systems near equilibrium,
when $\varphi_c$ is not expected to change
considerably with time. (This has been called the quasiadiabatic
approximation in Refs. \cite{morikawa}
and \cite{morikawa2}.)
We then obtain,

\begin{eqnarray}
\lefteqn{\frac{\lambda^2}{2} \varphi_c (x) \int d^3 x' \int_{-\infty}^t d t'
\varphi_c^2 (\vec{x'},t') {\rm Im} \left[G_\phi^{++}\right]_{x,x'}^2 \simeq}
\hspace{4cm}\nonumber \\
& & \simeq  \lambda^2 \varphi_c^2 (\vec x,t) \dot{\varphi}_c (\vec x,t)
\int_{-\infty}^{t} d t' (t' - t) \int \frac{d^3 q}{(2 \pi)^3} {\rm Im}
\left[G_\phi^{++}(\vec q , t-t')\right]^2 + \nonumber \\
& & + \frac{\lambda^2}{2} \varphi_c^3 (\vec x,t) \int_{-\infty}^{t} d t'
\int \frac{d^3q}{(2\pi)^3}
{\rm Im}\left[G_\phi^{++}(\vec q , t-t')\right]^2  \: .
\label{Diss1 approx}
\end{eqnarray}

The emergence of a time direction within this approximation is surely
related to neglecting the faster moving modes in the description of the
dynamics. This is an interesting question which deserves further
study, but that we will not address in the present work.
The last term in the left hand side of Eq. (\ref{full eqmotion}) can
also be worked out as in (\ref{Diss 1}) and
(\ref{Diss1 approx}) and we obtain

\begin{eqnarray}
\lefteqn{\frac{\lambda^2}{3}\int d^3 x' \int_{-\infty}^t d t'
\varphi_c (\vec{x'},t') {\rm Im}\left[G_\phi^{++}\right]_{x,x'}^3 \simeq}
\nonumber \\
& & \simeq \frac{\lambda^2}{3} \dot{\varphi}_c (\vec x,t)
\int_{-\infty}^{t} d t' (t' - t) {\rm Im} \left[\prod_{j=1}^3 \int
\frac{d^3 q_j}{(2 \pi)^3}G_\phi^{++}(\vec{q}_j , t-t')\right]
\delta(\vec q_1 + \vec q_2 + \vec q_3)
+ \nonumber \\
& & + \frac{\lambda^2}{3} \varphi_c (\vec x,t) \int_{-\infty}^t d t'
{\rm Im} \left[\prod_{j=1}^3 \int \frac{d^3 q_j}{(2 \pi)^3}
G_\phi^{++}(\vec{q}_j , t-t')\right] \delta(\vec q_1 + \vec q_2 + \vec q_3)
\: .
\label{Diss2 approx}
\end{eqnarray}

The first term in the rhs of (\ref{Diss1 approx}) and (\ref{Diss2 approx})
are the corresponding dissipative terms associated with the
fluctuation fields $\xi_1$ and $\xi_2$, respectively. The last
term in the rhs of (\ref{Diss1 approx}) is the one-loop
finite temperature correction
to the vertex [second graph in (\ref{action 2l})], while the last term in
the rhs of (\ref{Diss2 approx})
is the contribution to the finite temperature two-loop correction
to the mass coming from the ``setting sun'' diagram [the last
graph in (\ref{action 2l})]. The time integrations in
(\ref{Diss1 approx}) and (\ref{Diss2 approx}) can be easily performed by
using the expression for $G_\phi^{++}(\vec q, t-t')$ given in
(\ref{G of k}), and by changing the time integration variable to
$t - t' = t''$. However, if when computing the above dissipation
terms we use the free propagator expressions given in
(\ref{G><}), we would find that they both vanish.
We would also obtain the wrong results for
the temperature dependent vertex and mass corrections coming
from the second and fourth graphs in (\ref{action 2l}), as can be explicitly
checked.
The results would be quite different if, instead of free
propagators, we use dressed propagators.
In fact, consistency
demands that
we correct the propagators for the scalar field $\phi$ by the respective
self-energy contributions
at least up to two-loops and $O(\lambda^2)$, since this
is the order at which we are evaluating corrections
in perturbation theory. This situation is
analogous to the current resummation techniques being employed in the
improved versions of the electroweak effective potential, as can be seen
in Ref. \cite{ewpert}. We thus write the dressed propagator as

\begin{equation}
\frac{1}{q^2 - m^2 + i \epsilon} \rightarrow
\frac{1}{q^2 - m^2 - \Sigma(q) + i\epsilon} \: ,
\label{prop}
\end{equation}

\noindent
where $\Sigma(q)$ is the self-energy contribution,

\vspace{0.5cm}
\begin{equation}
\begin{picture}(300,28)
\put(0,0){$\Sigma(q) \: =$}
\thicklines
\put(60,0){\line(1,0){40}}
\put(80,10){\circle{20}}

\put(110,0){+}

\put(130,0){\line(1,0){40}}
\put(150,10){\circle{20}}
\put(150,30){\circle{20}}

\put(180,0){+}

\put(200,0){\line(1,0){40}}
\put(220,0){\circle{20}}

\put(250,0){+}

\put(270,0){${\cal O}(\lambda^3)\:\: .$}
\end{picture}
\label{self}
\end{equation}

\noindent
In Appendix A we show that the physical propagator $G_\phi^{++}(\vec q,
t-t')$ is then changed to

\begin{eqnarray}
G_\phi^{++}(\vec q ,t-t') & \simeq & \frac{e^{-\Gamma(\vec q) |t-t'|}}
{2 \omega(\vec q)} \left [ (1 + 2n) \cos \left [\omega |t-t'| \right ] -
i \sin \left [\omega|t-t'|\right ] + \right. \nonumber \\
&+& \left. 2 \beta \Gamma(\vec q) n(1 + n) \sin \left [\omega|t-t'|\right ]
+ {\cal O}\left (\frac{\Gamma^2}{T^2}\right ) \right ] \: ,
\label{full prop}
\end{eqnarray}

\noindent
where $\Gamma(\vec q)$ is the particle decay width \cite{weert}

\begin{equation}
\Gamma(q) = - \frac{{\rm Im}\Sigma(q)}{2 \omega(q)}
\label{def width}
\end{equation}

\noindent
and in (\ref{full prop}) we used the approximation $\beta \Gamma \ll 1$
(see appendix A), which is consistent with slow relaxation time-scales.
In (\ref{full prop}), $\omega\equiv\omega(\vec q)$ and
$n(\omega)$ are now given in terms of the finite temperature effetive
mass $m_T$

\begin{equation}
m_T^2 = m^2 + {\rm Re}\Sigma(m_T) \stackrel{T \gg m_T}{\simeq}
m^2 + \lambda \frac{T^2}{24} - \frac{\lambda^2 T^3}{384 \pi m_T} +
\frac{\lambda^2 T^2}{192 \pi^2} \ln \left( \frac{m_T^2}{T^2} \right) +
\ldots \:,
\label{mT}
\end{equation}

\noindent
where we have written explicitly the main thermal contributions from
each of the
terms in (\ref{self}). The second and third terms in the rhs of (\ref{mT})
are easily obtained. The last term in the rhs of (\ref{mT}), associated with
the ``setting sun'' diagram, is
explicitly evaluated in \cite{parwani}.

Using the dressed propagator (\ref{full prop}) in the expression for the
dissipation term (\ref{Diss1 approx}) and performing the
integration in $t'$, we obtain, to order $\lambda^2$,

\begin{eqnarray}
\lefteqn{\frac{\lambda^2}{2} \varphi_c (x) \int d^3 x' \int_{-\infty}^t
d t' \varphi_c^2 (\vec x,t') {\rm Im}\left[G_\phi^{++}\right]_{x,x'}^2
\simeq } \hspace{3.5cm}\nonumber \\
& & \simeq \frac{\lambda^2}{8} \varphi_c^2 (\vec x, t)
\dot{\varphi}_c (\vec x,t) \; \beta \int \frac{d^3q}{(2 \pi)^3}
\frac{n(1 + n)}{\omega^2 (\vec q) \Gamma(\vec q)} -\nonumber \\
& & - \frac{\lambda^2}{2} \varphi_c^3 (\vec x,t) \int \frac{d^3 q}
{(2 \pi)^3} \frac{1}{4 \omega^2 (\vec q)}\left[ \frac{1 + 2 n}{2
\omega(\vec q)} + \beta n(1 + n) \right] +
{\cal O}\left(\lambda^2 \frac{\Gamma}{\omega} \right) \: .
\label{diss}
\end{eqnarray}

\noindent
The first term in the rhs of (\ref{diss}) gives the dissipation term,
$\eta_1 \varphi_c^2 \dot{\varphi}_c$, with dissipation coefficient
$\eta_1$ given by

\begin{equation}
\eta_1 = \frac{\lambda^2}{8} \beta \int \frac{d^3 q}{(2 \pi)^3}
\frac{n(\omega)\left[1 + n(\omega)\right]}{\omega^2 (\vec q)
\Gamma (\vec q)} + {\cal O}\left(\lambda^2 \frac{\Gamma}{\omega}\right)\: .
\label{friction 1}
\end{equation}

\noindent
The second term in the rhs in (\ref{diss}), clearly gives just the
one-loop finite temperature vertex correction. In order to obtain
(\ref{diss}) we have performed an expansion to first order
in powers of $\Gamma/\omega$,
consistent with slowly varying modes. Also, since
$\Gamma \sim {\cal O}(\lambda^2)$, we have omitted the ${\cal O}(\lambda^4)$
contributions.
The expression for the dissipation coefficient can be further
simplified if we consider the high temperature limit $T\gg m_T$.
As shown in Refs. \cite{hosoya1,parwani} the high temperature limit of
$\Gamma(\vec q)$ is

\begin{equation}
\Gamma \stackrel{T \gg m_T}{\simeq} \frac{\lambda^2 T^2}{1536 \pi
\omega(\vec q)} \:.
\label{decay}
\end{equation}

Using (\ref{decay}) in (\ref{friction 1}), we obtain for $\eta_1$, in the
high temperature limit,

\begin{equation}
\eta_1 \stackrel{T \gg m_T}{\simeq} \frac{96}{\pi T} \ln \left(\frac{T}{m_T}
\right) \: ,
\label{high T fric}
\end{equation}

\noindent
which shows that the dissipation coefficient associated with the multiplicative
noise field $\xi_1$ is, in this limit,
only weakly (logarithmically) dependent on the
coupling constant $\lambda$.

We can  proceed in an analogous way and
evaluate Eq. (\ref{Diss2 approx}) in order to obtain
the expression for the dissipation coefficient associated with the second
fluctuation (noise) field $\xi_2$, from the first term in the rhs of
(\ref{Diss2 approx}). From the second term we can obtain the
two-loop mass correction coming from the fourth graph in (\ref{action 2l}).
Substituting Eq. (\ref{full prop}) for $G_\phi^{++}(\vec q_j,t-t')$ in
(\ref{Diss2 approx}) and performing the integration in $t'$, it is
possible to show (see appendix B) that the dissipation coefficient
associated with $\xi_2$ is at least of order $\lambda^2 \Gamma(\vec q_j)
\sim {\cal O}(\lambda^4)$. Therefore, in a weakly interacting
model, the dominant contribution to dissipation in the equation
of motion for $\varphi_c$  comes from the dissipation
term associated with the multiplicative noise field, $\xi_1$.

\subsection{The effective Langevin-like equation of motion}

Hence, up to two loops and
$O(\lambda^2)$, at zero external momentum and
within the adiabatic approximation
we obtain, from (\ref{full eqmotion}), the following
equation of motion for $\varphi_c$:

\begin{equation}
\left[ \Box + m_T^2 \right] \varphi_c (\vec x,t) +
\frac{\lambda_T}{3 !} \varphi_c^3 (\vec x,t) + \eta_1 \varphi_c^2 (\vec x,t)
\dot{\varphi}_c (\vec x , t) = \varphi_c (\vec x,t) \xi_1 (\vec x,t) \: ,
\label{eq motion}
\end{equation}

\noindent
where $\eta_1$ is given by (\ref{high T fric}),
$m_T$ and $\lambda_T$ are the renormalized finite temperature mass
and coupling constant, respectively, obtained from the renormalized
effective action,
Eq. (\ref{Seff}). The renormalization of $S_{\rm eff}$ can be defined by
the usual introduction of counterterms in the initial Lagrangian,
Eq. (\ref{Lagr}), by writing  ${\cal L} \to {\cal L} + \delta {\cal L}$, where
$\delta {\cal L} = \frac{1}{2} {\cal Z} (\partial_\mu \phi)^2 -
\frac{1}{2} \delta m^2 \phi^2 - \frac{\delta \lambda}{4 !} \phi^4$, with
${\cal Z}$, $\delta m^2$ and $\delta \lambda$ being the wave-function, mass
and vertex renormalization counterterms, respectively.
$\delta \lambda$ cancels the logarithmic divergence of the one-loop
vertex correction, while ${\cal Z}$ and $\delta m^2$ renormalize the
self-energy contribution, Eq. (\ref{self}).
In the high temperature limit, $m_T$ is given by Eq. (\ref{mT}) and $\lambda_T$
is given by

\begin{equation}
\lambda_T  \simeq \lambda - \frac{3 \lambda^2}{2} \left[
\frac{T}{8 \pi m_T} +
\frac{1}{8 \pi^2} \left[\ln\left(\frac{m_T}{4 \pi T} \right) + \gamma \right]
+ {\cal O}\left(\frac{m_T}{T}\right) \right]\: .
\label{Tparameters}
\end{equation}

\noindent
Eq. (\ref{eq motion}) can also be written in terms of a finite temperature
effective potential $V_{\rm eff} (\varphi_c,T)$,

\begin{equation}
\Box \varphi_c  + V'_{\rm eff}(\varphi_c,T) + \frac{96}{\lambda^2 \pi T}
\ln \left( \frac{T}{m_T}\right)
\left[V^{(3)}(\varphi_c)\right]^2  \dot{\varphi_c} =
\varphi_c \xi_1 \: ,
\label{eff motion}
\end{equation}

\noindent
where $V^{(3)}(\varphi_c) = \frac{d^3 V[\phi]}{d \phi^3}|_{\varphi_c}$.

Note that this equation, apart from the
important multiplicative noise source on the rhs, is analogous to
the one obtained
by Hosoya and Sakagami, using quite different methods,
for the evolution of the thermal average of the
scalar field $\varphi_c$ \cite{hosoya1}.

{}From the equation for the probability distribution for the
fluctuation field $\xi_1$, $P[\xi_1]$, Eq. (\ref{P1}), we have that
the two-point correlation function for $\xi_1 (x)$ is given by

\begin{equation}
\langle \xi_1 (x) \xi_1 (x')\rangle = \frac{\lambda^2}{2}
{\rm Re}\left[G_\phi^{++}\right]_{x,x'}^2 \: .
\label{corr noise}
\end{equation}

\noindent
Using (\ref{G2}) and (\ref{full prop}), we obtain for the two-point
correlation function (\ref{corr noise}) the expression (at zero
external momentum)

\begin{eqnarray}
\langle \xi_1 (x) \xi_1 (x')\rangle &=& \frac{\lambda^2}{2}
\delta^3(\vec x - \vec{x'}) \int \frac{d^3 q}{(2 \pi)^3}
\frac{1}{4 \omega^2(\vec q)} \left \{ 2 n(\omega) \left [1 + n(\omega)
\right ] +
\right. \nonumber \\
&+& \left. [1 + 2 n(\omega) + 2 n^2 (\omega)] \cos\left [2 \omega|t-t'|
\right ] + \right.
\nonumber \\
& + & \left. 2 \beta \Gamma(\vec q) n(\omega)[1 + n(\omega)][1 + 2 n(\omega)]
\sin[2 \omega |t -t'|] \right \} e^{-2 \Gamma(\vec q) |t - t'|}
+ {\cal O}\left( \lambda^2 \frac{\Gamma^2}{T^2}\right) \: ,
\label{corr 1}
\end{eqnarray}

\noindent
which shows that the noise is colored (time dependent), although it
is Gaussian distributed.
Up to order $\lambda^2$ and
for $\Gamma/\omega \ll 1 \: , \: \Gamma/T \ll 1$, we obtain the
fluctuation-dissipation relation

\begin{equation}
\eta_1 = \frac{1}{T} \int d^4 x' \langle \xi_1 (x) \xi_1 (x') \rangle
\theta(t-t') \:.
\label{fluc-diss}
\end{equation}

\noindent
We can also obtain the Markovian limit of (\ref{corr 1}),
that is, the limit in which the noise is uncorrelated (white).
Note that as $T \to \infty$, $\Gamma \to
\infty$, and thus the integrand becomes sharply peaked at $|t-t^{\prime}|\sim
0$. In this limit, we can approximate
(\ref{corr 1}) by

\begin{eqnarray}
\langle \xi_1 (x) \xi_1 (x') \rangle \stackrel{T \to \infty}{\longrightarrow}
& \frac{\lambda^2}{2}&  \delta(\vec x - \vec x') \delta(t -t')
\int \frac{d^3 q}{(2 \pi)^3} \frac{n(\omega)[1 +n(\omega)]}{2 \omega^2
(\vec q \;) \Gamma(\vec q \;)}= \nonumber \\
&=& 2 T \eta_1 \delta(\vec x - \vec x') \delta(t -t')  \: ,
\label{white}
\end{eqnarray}

\noindent
where $\eta_1$ is given by (\ref{friction 1}). Eq. (\ref{white}) is the
standard expression of the fluctuation-dissipation theorem for a
Gaussian white noise.

\section{Coupling the Scalar Field to other Fields}

The previous computation of the effective equation of motion
for the field configuration $\varphi_c$  can be generalized to include
the effects of interactions with other fields.
As an example, consider the
Lagrangian density for the scalar field $\phi$ interacting quadratically
with
another scalar field $\chi$,

\begin{equation}
{\cal L}[\phi,\chi] = \frac{1}{2} (\partial_\mu \phi)^2 +\frac{1}{2}
(\partial_\mu \chi)^2 - V[\phi,\chi] \: ,
\label{L 2fields}
\end{equation}

\noindent
with potential $V[\phi,\chi]$ given by

\begin{equation}
V[\phi,\chi] = \frac{m^2}{2}\phi^2 + \frac{\lambda}{4 !} \phi^4
+ \frac{\mu^2}{2}\chi^2 + \frac{f}{4!} \chi^4 + \frac{g^2}{2} \phi^2 \chi^2 \:,
\label{V 2fields}
\end{equation}

\noindent
where $m^2$ and $\mu^2$ are positive. This model is a good toy model
for several physical cases of interest. For example, for some
relations among the values of the coupling constants $\lambda$, $f$ and
$g^2$ (e.g. $\lambda \sim {\cal O}(g^4)$, $f \sim {\cal O}(g^2)$
\cite{weinberg}), Eq. (\ref{L 2fields}) exhibits the properties of
Coleman-Weinberg models, for which the quantum corrections coming from
integrating out the $\chi$ field break the symmetry in the potential
for the scalar field $\phi$ (corrected by the $\chi$-loop quantum
corrections), modifying the original vacuum structure of
the model.
Also, as pointed out by Hu, Paz and Zhang \cite{hu1}, (\ref{L 2fields}) can
mimic, at lowest order (one-loop) in the
$\chi$-loop quantum corrections, a coarse-grained effective model for
the scalar field $\phi$, after integrating out the $\chi$ field. In this
case, the $\phi$ field would represent the field with components
containing the long wavelength modes, while $\chi$ would contain the
short wavelength modes, with a cutoff
determined by some scale $\Lambda$. In
inflationary models, $\phi$ would behave as a classical
field, while $\chi$ would represent the sub-horizon high frequency
modes
 \cite{staro,rey,hu1}.
The authors in
\cite{hu1} thus consider the field $\chi$ as the quantum bath (at $T=0$),
allowing them to obtain an effective action
for the scalar field $\phi$ (the classical action corrected by
the $\chi$ field one-loop quantum corrections), where the scalar field
is coupled to a noise field, very much like
the multiplicative noise field $\xi_1$ in
Eq. (\ref{Seff}). {}Following the results of the last section, the
generalization of their results to $T\neq0$ is relatively simple.
Up to one-loop
in the $\chi$ field, the effective
action $\Gamma[\varphi]$ in (\ref{action 2l}) (also called the influence
functional by some authors),
will be given by

\begin{equation}
\Gamma[\varphi] \rightarrow \Gamma[\varphi] +
\frac{1}{2} i {\rm Tr}_c \ln \left[ \Box + \mu^2 + g^2 \varphi^2 \right] \:.
\label{action 2fields}
\end{equation}

\noindent
Expanding the logarithm in (\ref{action 2fields}) as in (\ref{Trlnsum}),
up to order $g^4$, we will get expressions analogous to the ones
in (\ref{1loop1}) and (\ref{1loop2}), with $\varphi_{+}$, $\varphi_{-}$
in the external legs and internal propagators
for the $\chi$ field, $G_{\chi}^{n,l}(x -x')$ ($n,l = +,-$),
with expressions just as in (\ref{Gmatrix}) and (\ref{G of k}).
By changing the field variables $\varphi_{+}$, $\varphi_{-}$ to
$\varphi_{\Delta}$, $\varphi_c$ as before, the contribution from the $\chi$
field to the effective action for $\phi$,
$\Gamma[\varphi_{\Delta},\varphi_c]$, Eq. (\ref{full action}), will be

\begin{eqnarray}
\lefteqn{\Gamma[\varphi_\Delta,\varphi_c] \rightarrow \Gamma[\varphi_\Delta,
\varphi_c] - g^2 \int d^4 x \varphi_\Delta (x) \varphi_c (x) \int
\frac{d^3 k}{(2 \pi)^3} \frac{1 + 2 n_\chi}{\omega_\chi} -}
\nonumber \\
& & - \frac{g^4}{2} \int d^4 x d^4 x' \left[ \varphi_\Delta (x)
\varphi_c (x) \varphi_\Delta^2 (x') + 4 \varphi_\Delta (x) \varphi_c (x)
\varphi_c^2 (x') \right] {\rm Im} \left[G_\chi^{++}\right]_{x,x'}^2
\theta(t-t') +\nonumber \\
& & + i g^4 \int d^4 x d^4 x' \varphi_\Delta (x) \varphi_c (x)
\varphi_\Delta (x') \varphi_c (x') {\rm Re}\left[G_\chi^{++}\right]_{x,x'}^2
\: ,
\label{full action2}
\end{eqnarray}

\noindent
where $\left[G_\chi^{++}\right]_{x,x'}^2$ is given by an expression analogous
to (\ref{G2}).

The imaginary term in (\ref{full action2}), coming from integrating out
the $\chi$ field (at one-loop order)
can be rewritten by redefining the fluctuation field $\xi_1$ in
$(\ref{Seff})$, such that its probability distribution in (\ref{P1}) is
changed to

\begin{equation}
P[\xi_1] = N_1^{-1} \exp\left\{ - \frac{1}{2} \int d^4 x d^4 x'
\xi_1 (x) \left[ \frac{\lambda^2}{2} {\rm Re} \left[G_\phi^{++}\right]_{x,x'}^2
+ 2 g^4 {\rm Re} \left[G_\chi^{++}\right]_{x,x'}^2 \right] \xi_1 (x')
\right\} \: .
\label{P1 new}
\end{equation}

\noindent
and the two-point correlation function for $\xi_1$ is now given by

\begin{equation}
\langle \xi_1 (x) \xi_1 (x') \rangle = \frac{\lambda^2}{2}
{\rm Re} \left[G_\phi^{++}\right]_{x,x'}^2
+ 2 g^4 {\rm Re} \left[G_\chi^{++}\right]_{x,x'}^2  \: .
\label{corr new}
\end{equation}

\noindent
In the equation of motion for $\varphi_c$, from (\ref{full action2}), we
will have an additional contribution to the dissipation coefficient
$\eta_1$, obtained from a term analogous to (\ref{Diss1 approx}):

\begin{eqnarray}
\lefteqn{2 g^4 \int d^4 x' \varphi_c^2 (x') {\rm Im}
\left[G_\chi^{++}\right]_{x,x'}^2  \theta(t-t') \simeq}
\hspace{4cm}\nonumber \\
& & \simeq 4 g^4 \varphi_c^2 (\vec x, t) \dot{\varphi}_c (\vec x ,t)
\int_{-\infty}^t d t' (t'-t) \int \frac{d^3 q}{(2 \pi)^3}
{\rm Im} \left[G_\chi^{++}(\vec q,t-t')\right]^2 + \nonumber \\
& & +2 g^4 \varphi_c^3 (\vec x,t) \int_{-\infty}^t d t' \int
\frac{d^3 q}{(2 \pi)^3}
{\rm Im}
\left[G_\chi^{++}(\vec q,t-t')\right]^2 \:.
\label{new diss}
\end{eqnarray}

\noindent
The first term in the rhs in (\ref{new diss}) gives the contribution to
the dissipation coefficient $\eta_1$, due to the interaction of the scalar
field $\phi$ with the $\chi$ field. The second term in (\ref{new diss}),
together with the second term in the rhs in (\ref{full action2}),
give the corrections of order $g^4$ and $g^2$ to the scalar field $\phi$
vertex and mass, respectively, due to the $\chi$-loop quantum corrections.

As in (\ref{Diss1 approx}), in order to obtain a nonvanishing
contribution to the dissipation
coefficient coming from (\ref{new diss}),
we must consider the
dressed propagator $G_\chi^{++}$ for the $\chi$ field, instead of the free
propagator. $G_\chi^{++}(\vec q ,t-t')$ has an expression similar to the
one given for the scalar field $\phi$, Eq. (\ref{Afull prop2})
(or (\ref{full prop}), for $\beta \Gamma_\chi \ll 1$), where, at
${\cal O}$-($g^4$, $f^2$), the decay width $\Gamma_\chi$
can be written as,
\begin{equation}
\Gamma_\chi (q) = - \frac{{\rm Im} \Sigma_\chi (\vec q, \omega_\chi)}
{2 \omega_\chi (q)} \: ,
\label{decay2}
\end{equation}

\noindent
with the imaginary part of the $\chi$-field self-energy, from (\ref{Im}),
given by the imaginary part of the two two-loop contributions below,

\begin{eqnarray}
\begin{picture}(330,28)
\put(-65,0){${\rm Im}\Sigma\chi\; = $}

\thicklines
\put(-10,0){${\rm Im} \;\left[ \hspace{3cm} \right]\;+\;
{\rm Im}\; \left[ \hspace{3cm} \right]\;=$}
\multiput(25,0)(8,0){8}{\line(1,0){4}}
\put(55,0){\circle{25}}

\put(50,-23){$\phi$}
\put(50,17){$\phi$}

\multiput(160,0)(8,0){8}{\line(1,0){4}}
\put(180,0){\line(0,1){1}}
\multiput(180,1)(0.04,0.2){5}{\line(1,0){0.1}}
\multiput(182.5,7.1)(-0.1,-0.15){6}{\line(1,0){0.1}}
\multiput(182.5,7.1)(0.1,0.1){14}{\line(1,0){0.1}}
\multiput(183.9,8.5)(0.15,0.1){6}{\line(1,0){0.1}}
\multiput(190,11)(-0.2,-0.04){5}{\line(1,0){0.1}}
\put(190,11){\line(1,0){2}}
\multiput(192,11)(0.2,-0.04){5}{\line(1,0){0.1}}
\multiput(199.5,7.1)(0.1,-0.15){6}{\line(1,0){0.1}}
\multiput(199.5,7.1)(-0.1,0.1){14}{\line(1,0){0.1}}
\multiput(198.1,8.5)(-0.15,0.1){6}{\line(1,0){0.1}}
\multiput(202,1)(-0.04,0.2){5}{\line(1,0){0.1}}
\put(202,0){\line(0,1){1}}
\put(180,0){\line(0,-1){1}}
\multiput(180,-1)(0.04,-0.2){5}{\line(1,0){0.1}}
\multiput(182.5,-7.1)(-0.1,0.15){6}{\line(1,0){0.1}}
\multiput(182.5,-7.1)(0.1,-0.1){14}{\line(1,0){0.1}}
\multiput(183.9,-8.5)(0.15,-0.1){6}{\line(1,0){0.1}}
\multiput(190,-11)(-0.2,0.04){5}{\line(1,0){0.1}}
\put(190,-11){\line(1,0){2}}
\multiput(192,-11)(0.2,0.04){5}{\line(1,0){0.1}}
\multiput(199.5,-7.1)(0.1,0.15){6}{\line(1,0){0.1}}
\multiput(199.5,-7.1)(-0.1,-0.1){14}{\line(1,0){0.1}}
\multiput(198.1,-8.5)(-0.15,-0.1){6}{\line(1,0){0.1}}
\multiput(202,-1)(-0.04,-0.2){5}{\line(1,0){0.1}}
\put(202,0){\line(0,-1){1}}

\put(187,-23){$\chi$}
\put(187,17){$\chi$}

\end{picture}
\nonumber
\end{eqnarray}

\vspace{0.2cm}

\begin{eqnarray}
= &-& \left( 1- e^{-\beta q_0} \right) \left[ \prod_{j=1}^{3}
\int \frac{d^4 k_j}{(2 \pi)^4} \left[1 + n(k_j^0)\right] \right]
\times \nonumber \\
& \times &
\left(g^4 \rho_\phi (k_1) \rho_\chi (k_2) \rho_\phi (k_3)
+ \frac{f^2}{12} \rho_\chi (k_1) \rho_\chi (k_2) \rho_\chi (k_3)\right)
(2 \pi)^4 \delta^4 (q - k_1-k_2-k_3)\:,
\label{Im X}
\end{eqnarray}

\noindent
where $\rho_\phi (k)$ is given by (\ref{spectral}), with $m^2$ and $\lambda$
corrected by the $\chi$-loop ($T\neq0$) quantum corrections.
$\rho_\chi (k)$ is the spectral function for the scalar field $\chi$,
with expression analogous to the one for the scalar field $\phi$, given by
Eq. (\ref{spectral}), but now with $\Gamma_\chi$ given by (\ref{decay2}) and
$\omega_\chi$
given by the solution of $\omega_\chi^2 (q) = \vec{q\;}^2 + \mu^2 +
{\rm Re}\Sigma_\chi (\vec q, \omega_\chi)$.

The high temperature limit of (\ref{decay2}) is analogous to the one
for the case with one scalar field $\phi$, Eq. (\ref{decay}),

\begin{equation}
\Gamma_\chi (\vec q) \stackrel{T \gg \mu_T}{\simeq}
\frac{T^2}{128 \pi \omega_\chi (\vec q)} \left(g^4 + \frac{f^2}{12} \right)\: ,
\label{decay2 highT}
\end{equation}

\noindent
with $\mu_T^2 = \mu^2 + {\rm Re}\Sigma_\chi (\mu_T)$. Using these in
(\ref{new diss}), we obtain an equation of motion for
$\varphi_c$ still written as in (\ref{eq motion}), up to
two loops and order $\lambda^2$ in the scalar field $\phi$ and up to
one-loop\footnote{Up to 2 loops
in $\chi$ the situation would be identical as discussed for the scalar
field $\phi$, with results similar to the last section and that of
appendix B.} and order $g^4$ in the scalar field $\chi$. The
dissipation coefficient $\eta_1$ is given by

\begin{eqnarray}
\eta_1 &=& \frac{\lambda^2}{8} \beta \int \frac{d^3 q}{(2 \pi)^3}
\frac{n_\phi (1 + n_\phi)}{\omega_\phi^2 (\vec q) \Gamma_\phi (\vec q)}
+ \frac{g^4}{2} \beta \int \frac{d^3 q}{(2 \pi)^3}
\frac{n_\chi (1 + n_\chi)}{\omega_\chi^2 (\vec q) \Gamma_\chi (\vec q)} +
{\cal O}\left(\lambda^2 \frac{\Gamma_\phi}{\omega_\phi}\right) +
{\cal O} \left(g^4 \frac{\Gamma_\chi}{\omega_\chi}\right) \nonumber \\
& \simeq & \frac{96}{\pi T} \left[ \ln \left( \frac{T}{m_T}\right) +
\frac{4 g^4}{12 g^4 + f^2} \ln \left( \frac{T}{\mu_T}\right) \right] \: ,
\label{new fric}
\end{eqnarray}

\noindent
with the second correction for $\eta_1$ coming from the $\chi$-$\phi$
interaction in (\ref{L 2fields}). Associated with this modified
dissipation term there is a modified
multiplicative fluctuation (noise) field $\xi_1$, with probability
distribution given by (\ref{P1 new}).
{}For a Coleman-Weinberg potential, we have that
$\lambda \sim {\cal O}(g^4)$ and
$f \sim {\cal O}(g^2)$, so that the dissipation
coefficient is, as in (\ref{high T fric}),
weakly dependent on the coupling constants within our
approximations.
Using the expression for the
two-point correlation function for $\xi_1$, Eq. (\ref{corr new}), both the
fluctuation-dissipation relation, Eq. (\ref{fluc-diss}) and
the Markovian limit expression, Eq. (\ref{white}),  still hold.

\section{Conclusions}

In this work we have studied the nonequilibrium dynamics of a self-coupled
scalar field. Even though our formalism is in principle applicable in
situations far from equilibrium, the effective Langevin-like equation we
obtained is only adequate to study the approach to equilibrium if the
initial conditions are not too far from equilibrium. This limitation
is essentially due to the use of perturbation theory and should come as no
surprise. However, this approach clarifies many important issues concerning
nonequilibrium fields and the nature of the system-bath coupling. By
integrating over fluctuations in order to obtain the effective action, it
becomes clear in what sense the short wavelength modes can function as the
thermal bath that drives the longer wavelength modes into equilibrium.
In this sense, the approximations employed in order to obtain a
Langevin-like equation are consistent with this system-bath separation;
longer wavelength modes have slower dynamics and are responsible for the
large-distance coherent behavior observed during the approach to equilibrium
both in the laboratory and in numerical simulations. By going to higher
order in perturbation theory we were able to obtain the
contributions to the noise and dissipation terms coming from different
diagrams and their relevance to the nonequilibrium dynamics.

We found that the Langevin-like equation describing the approach to
equilibrium both for a self-coupled scalar field and for
quadratic coupling with other fields
is quite different from the usual phenomenological form
with Gaussian white noise used so far in numerical
simulations of the approach to equilibrium in field theory.
There are basically three differences. The first is that the dominant
contribution to the noise is
multiplicative; it couples quadratically to the field, acting as a ``noisy''
source to the mass term in the equation of motion. The second
difference is that even though this noise is still
Gaussian distributed, it is now  non-Markovian; the correlation times
depend on the decay width of the fluctuations generating the noise. As we
show in the text, only in the limit of very high temperatures the noise
becomes white, as one would naively expect. The final difference has to do with
the way the dissipation term appears in the equation of motion. Instead
of the simple $\eta \dot {\phi}$ term, we find instead the dissipation
``coefficient'' depends quadratically on the amplitude of the field,
$\eta(T) \phi^2 \dot {\phi}$. In the high temperature limit for a single
scalar field we obtained
that $\eta(T)\sim (1/T){\rm ln}(T/m_T)$, being thus weakly dependent on the
coupling constant.
This result is in agreement with the work of
Ref. \cite{hosoya1}, which assumed a small departure from equilibrium
within a kinetic approach. Both results are consistent with
linear response theory commonly used to obtain transport coefficients in
field theories \cite{jeon}.

By studying the effects of another scalar field quadratically coupled
to $\phi$ we were able to obtain their different contributions to the
noise and dissipation terms in the effective equation of motion. Now, the
coefficient of the dissipation term depends on ratios of couplings, as one
would expect in more realistic situations, while the
noise is still Gaussian and multiplicative. In both cases we showed
that one can recover a fluctuation-dissipation relation.
It will be interesting to investigate the implications of this Langevin-like
equation to the equilibration time-scales during phase transitions,
by employing it in numerical simulations.
Apart from studying the approach to equilibrium from  near-equilibirum initial
conditions, it is possible to use this equation in the study of finite
temperature symmetry restoration, if one takes into account the effects
of expanding about the broken-symmetric vacuum. In this connection,
it is interesting to note that the coefficient of the dissipation term, in the
high-temperature limit, displays the typical critical slowing down
(poor infra-red behavior) observed
in many second-order phase transitions; since $\eta\sim {\rm ln}(T/m_T)$, as
the critical temperature is approached from below the temperature corrected
mass vanishes and the viscosity diverges logarithmically. We leave as an
open question the potential impact that a better understanding of
nonequilibrium dynamics of field theories will have on our current
modelling of primordial phase transitions and their possible observational
consequences.
However, we believe that interesting physics is lurking
behind our present level of understanding of nonequilibrium
physical processes that took
place in the early Universe.

\acknowledgments

We would like to thank M. Morikawa, E. Mottola, and S. Habib for useful
discussions.
(MG) was supported in part by a National Science
Foundation grant No. PHYS-9204726. (ROR) is supported by a postdoctoral
grant from
Conselho Nacional
de Desenvolvimento Cient\'{\i}fico e Tecnol\'ogico - CNPq (Brazil).
\appendix

\section{}

In this Appendix we obtain the expression for the dressed
scalar field propagator,
Eq. (\ref{full prop}).
The finite temperature, real-time propagator $G_\phi^{++}
(\vec q,t-t')$, can be written
in terms of the spectral function $\rho(\vec q,q_0)$
\cite{weert,jeon},

\begin{equation}
G_\phi^{++}(\vec q,t-t') = \int_{-\infty}^{+\infty} \frac{d q_0}{2 \pi}
\rho(\vec q,q_0) \left \{ \left [ 1 + n(q_0)\right ] \theta(t-t') +
n(q_0) \theta(t'-t) \right \} \: ,
\label{Aprop}
\end{equation}

\noindent
where $n(q_0) = \frac{1}{e^{\beta q_0} - 1}$ and the spectral function, for
the dressed propagator (\ref{prop}), is

\begin{equation}
\rho(\vec q,q_0) = i \left[ \frac{1}{(q_0 + i \Gamma)^2 - \omega^2 (q)} -
\frac{1}{(q_0 - i \Gamma)^2 - \omega^2(q)} \right] \: ,
\label{spectral}
\end{equation}

\noindent
where $\omega(q)$ is the solution of $\omega^2 (q) = \vec{q}^2 + m^2 +
{\rm Re}\Sigma (\vec q,\omega)$ and $\Sigma(q)$ is the scalar field
self-energy, given by (\ref{self}), up to two loops. The spectral
function (\ref{spectral}) has a peak at $q_0 = \omega(q)$, with a
width given by $\Gamma \equiv \Gamma(q)$,

\begin{equation}
\Gamma(q) = - \frac{{\rm Im}\Sigma(\vec q,\omega)}{2 \omega(q)} \:.
\label{Awidth}
\end{equation}

{}For the free propagator,

\begin{equation}
\rho(\vec q,q_0) = i \left[ \frac{1}{(q_0 + i \epsilon)^2 -
\vec{q}^2 - m^2} -   \frac{1}{(q_0 - i \epsilon)^2 -
\vec{q}^2 - m^2} \right]
\label{free spectral}
\end{equation}

\noindent
and $\rho(\vec q,q_0) \stackrel{\epsilon \to 0}{\longrightarrow} 2 \pi
\varepsilon(q_0) \delta(q^2 - m^2)$, where $\varepsilon(q_0) =
\theta(q_0) - \theta(-q_0)$. Substituting (\ref{free spectral}) in
(\ref{Aprop}), we obtain the free propagator expressions in
(\ref{G of k})-(\ref{G><}).

Eq. (\ref{spectral}) has four poles in the complex $q_0$ plane:
$\omega\pm i \Gamma$ and $-\omega \pm i\Gamma$. Using (\ref{spectral})
in (\ref{Aprop}) and performing the $q_0$ integration, we obtain

\begin{equation}
G_\phi^{++} (\vec q , t-t') = G_\phi^{>} (\vec q,t-t') \theta(t-t')
+ G_\phi^{<} (\vec q,t-t') \theta(t'-t) \: ,
\label{Afull prop}
\end{equation}

\noindent
where

\begin{eqnarray}
& G_\phi^{>}(\vec q, t-t') & = \frac{1}{2 \omega} \left\{ \left[
1+ n(\omega - i \Gamma)\right] e^{-i (\omega - i \Gamma) (t-t')} +
n(\omega + i \Gamma) e^{i (\omega + i \Gamma) (t-t')} \right\} \:,
\nonumber \\
& G_\phi^{<}(\vec q, t-t') & = G_\phi^{>}(\vec q,t'-t) \:.
\label{AG><}
\end{eqnarray}

\noindent
The expressions for $G_\phi^{--}$, $G_\phi^{+-}$ and $G_\phi^{-+}$ are
the same as in (\ref{G of k}), but with $G_\phi^{>,<}$ given now by
(\ref{AG><}).

$\Gamma(q)$ is given in terms of the imaginary part of the self-energy
(coming from the third graph in (\ref{self})) by (\ref{Awidth}), where
${\rm Im} \Sigma(q)$ is \cite{hosoya2,jeon}

\begin{eqnarray}
\begin{picture}(320,28)
\put(-70,0){${\rm Im}\Sigma (q)\; = $}

\thicklines
\put(-5,0){${\rm Im} \;\left[ \hspace{3cm} \right]\;=$}
\put(30,0){\line(1,0){60}}
\put(60,0){\circle{25}}
\end{picture}
\nonumber
\end{eqnarray}

\begin{equation}
= - \frac{\lambda^2}{12} \left(1 - e^{-\beta q_0}\right)
\left[ \prod_{j=1}^{3} \int \frac{d^4 k_j}{(2 \pi)^4} \rho (k_j)
\left[ 1 + n(k_j^0)\right] \right] (2 \pi)^4 \delta^4 (q - k_1-k_2-k_3)\:.
\hspace{2.5cm}
\label{Im}
\end{equation}

\noindent
The high temperature limit
of (\ref{Awidth}) is given in Refs. \cite{hosoya1} and
\cite{parwani} and we have just quoted the final result in the text.

The expression for $G_\phi^{++}(\vec q,t-t')$ in (\ref{Afull prop}) can
also be explicitly written as

\begin{eqnarray}
G_\phi^{++}(\vec q,t-t') &=& \frac{e^{-\Gamma |t-t'|}}{2 \omega
\left[ \cosh(\beta \omega) - \cos (\beta \Gamma)\right]}
\left \{ \sinh (\beta \omega) \cos (\omega |t-t'|) +\right.
\nonumber \\
&+& \left.\sin(\beta \Gamma) \sin(\omega |t-t'|) +
i \left[\cos(\beta\Gamma) - \cosh(\beta \omega)\right]
\sin(\omega |t-t'|) \right \} \:.
\label{Afull prop2}
\end{eqnarray}

\noindent
Expanding (\ref{Afull prop2}) for $\beta \Gamma \ll 1$, we obtain Eq.
(\ref{full prop}).

\section{}

We estimate here the dissipation coefficient $\eta_2$ associated with
the fluctuation field $\xi_2$, obtained
from Eq. (\ref{Diss2 approx}), with $G_\phi^{++}(\vec q ,t-t')$
given by (\ref{Afull prop2}) and show that
it is subdominant.
{}From the first term in the rhs in (\ref{Diss2 approx}), we get that
$\eta_2$ is given by

\begin{equation}
\eta_2 = \frac{\lambda^2}{3} \int_{-\infty}^t d t' (t'-t)
\int \frac{d^3 q_1}{(2 \pi)^3}\frac{d^3 q_2}{(2 \pi)^3}
{\rm Im} \left[ G_\phi^{++}(\vec{q}_1,t-t')G_\phi^{++}(\vec{q}_2,t-t')
G_\phi^{++}(-\vec{q}_1-\vec{q}_2,t-t') \right] \:.
\label{Bdiss}
\end{equation}

\noindent
{}From (\ref{Afull prop2}), we can write $G_\phi^{++}(\vec{q}_j,t-t')$ as

\begin{equation}
G_\phi^{++}(\vec{q}_j,t-t') = a_j + i b_j \: ,
\label{Gab}
\end{equation}

\noindent
where $a_j \equiv a(\vec{q}_j,t-t')$ and $b_j \equiv b(\vec{q}_j,t-t')$
are given by the real and imaginary terms of Eq. (\ref{Afull prop2}),
respectively ($\Gamma_j \equiv \Gamma(\vec{q}_j)$ and $\omega_j \equiv
\omega(\vec{q}_j)$) :

\begin{eqnarray}
& a_j & =
\frac{e^{-\Gamma_j |t-t'|}}{2 \omega_j
\left [ \cosh(\beta \omega_j) - \cos (\beta \Gamma_j)\right ]}
\left [ \sinh (\beta \omega_j) \cos (\omega_j |t-t'|)
+ \sin(\beta \Gamma_j) \sin(\omega_j |t-t'|) \right ] \:,
\nonumber \\
\nonumber \\
& b_j & = - \; \frac{e^{-\Gamma_j |t-t'|} \sin(\omega_j |t-t'|)}{2 \omega_j}\:.
\label{ab}
\end{eqnarray}

\noindent
Using (\ref{Gab}) in (\ref{Bdiss}), we get (with $\vec{q}_3 = -\vec{q}_1
-\vec{q}_2$)

\begin{equation}
\eta_2 = \frac{\lambda^2}{3} \int_{-\infty}^t d t' (t'-t)
\int \frac{d^3 q_1}{(2 \pi)^3}\frac{d^3 q_2}{(2 \pi)^3}
\left(a_1 a_2 b_3 + a_1 a_3 b_2 + a_2 a_3 b_1 -b_1 b_2 b_3 \right) \:.
\label{Bdiss2}
\end{equation}

\noindent
Using Eq. (\ref{ab}) for $a_j$ and $b_j$, we can perform the time integration
in (\ref{Bdiss2}), by changing the time integration variable $t'$ to
$t-t'=t''$, and obtain for (\ref{Bdiss2}) the expression

\begin{eqnarray}
\eta_2 &=& \frac{\lambda^2}{3}
\int \frac{d^3 q_1}{(2 \pi)^3}\frac{d^3 q_2}{(2 \pi)^3}
\frac{1}{\omega_1 \omega_2 \omega_3} \left\{
\frac{\Gamma_1 +\Gamma_2 +\Gamma_3}{2} \left[
\frac{2(1+2n_1)(n_2 -n_3) + (1+2n_2)(1 +2n_3) -1}{(\omega_1-
\omega_2 + \omega_3)^3} + \right.\right. \nonumber \\
&+& \left.\left. \frac{2(1 + 2n_2)(n_3-n_1) + (1 + 2n_1)(1+2n_3)-1}
{(\omega_1 +\omega_2 -\omega_3)^3} +\right.\right.\nonumber \\
&+& \left.\left.
\frac{2(1 + 2n_3)(n_1 -n_2) + (1 + 2n_1)(1 + 2n_2) -1}{(-\omega_1 +
\omega_2 + \omega_3)^3} -\right.\right. \nonumber \\
&-& \left.\left.\frac{(1 + 2n_1)(1 + 2n_2) + (1+ 2n_1)(1 +2n_3) +
(1+2n_2)(1+2n_3) -1}
{(\omega_1 +\omega_2+\omega_3)^3} \right] + \right.\nonumber \\
&+& \left. \beta \Gamma_1 n_1 (1 + n_1) \left[
(n_2-n_3) \left(\frac{1}{(\omega_1-\omega_2+\omega_3)^2}- \frac{1}{
(\omega_1+\omega_2-\omega_3)^2}\right) +\right.\right.\nonumber\\
&+& \left.\left.
(n_2 + n_3) \left(\frac{1}{(\omega_1+\omega_2+\omega_3)^2}- \frac{1}{
(-\omega_1+\omega_2+\omega_3)^2}\right) \right]+\right. \nonumber \\
&+& \left. \beta \Gamma_2 n_2 (1 + n_2) \left[
(n_1 - n_3) \left( \frac{1}{(-\omega_1+\omega_2+\omega_3)^2}- \frac{1}{
(\omega_1+\omega_2-\omega_3)^2}\right) +\right.\right.\nonumber \\
&+& \left.\left.
(n_1 + n_3) \left( \frac{1}{(\omega_1+\omega_2+\omega_3)^2}- \frac{1}{
(\omega_1-\omega_2+\omega_3)^2}\right) \right]+ \right.
\nonumber \\
&+& \left. \beta \Gamma_3 n_3 (1 + n_3) \left[
(n_1 - n_2) \left(\frac{1}{(-\omega_1+\omega_2+\omega_3)^2}- \frac{1}{
(\omega_1-\omega_2+\omega_3)^2}\right) +\right.\right.\nonumber \\
&+& \left.\left.
(n_1 + n_2) \left(\frac{1}{(\omega_1+\omega_2+\omega_3)^2}- \frac{1}{
(\omega_1-\omega_2+\omega_3)^2}\right) \right] \right\} +
{\cal O}\left(\lambda^2 \frac{\Gamma_j^3}{\omega_j^3}\right)\: .
\label{Big n2}
\end{eqnarray}

\noindent
The above expression is at least of order $\lambda^2 \Gamma_i$ and,
since $\Gamma_i \sim {\cal O}(\lambda^2)$, we have that
$\eta_2 \sim {\cal O}(\lambda^4)$. We are thus justified in
neglecting its contribution
to the effective equation of motion to ${\cal O}(\lambda^2)$.

\end{document}